\documentclass[%
 reprint,
 amsmath,amssymb,
 aps,
]{revtex4-2}

\usepackage{graphicx}
\usepackage{dcolumn}
\usepackage{bm}

\usepackage{color}
\begin{document}


\title{Constraining the position of the CEP through the Speed of Sound in the LSMq}

\author{Sa\'ul Hern\'andez-Ortiz}
 \email{saul.ortiz@umich.mx}
\affiliation{Instituto de F\'isica y Matem\'aticas, Universidad
Michoacana de San Nicol\'as de Hidalgo, Edificio C-3, Ciudad
Universitaria, Morelia, Michoac\'an 58040, M\'exico.}

\author{R. Mart\'inez von Dossow}%
 \email{ricardo.martinez@correo.nucleares.unam.mx}
\affiliation{Instituto de Ciencias Nucleares, Universidad Nacional Aut\'onoma de M\'exico, 04510 Ciudad de M\'exico, M\'exico.} %

%

\author{Alfredo Raya}
 \email{alfredo.raya@umich.mx}
\affiliation{Instituto de F\'isica y Matem\'aticas, Universidad
Michoacana de San Nicol\'as de Hidalgo, Edificio C-3, Ciudad
Universitaria, Morelia, Michoac\'an 58040, M\'exico,}
\affiliation{Centro de Ciencias Exactas, Universidad del B\'{\i}o-B\'{\i}o,\\ Avda. Andr\'es Bello 720, Casilla 447, 3800708, Chill\'an, Chile.}
%

\date{\today}

\begin{abstract}
We study the chiral phase transition within the Linear Sigma Model with quarks  from its thermodynamical potential considering quantum corrections up to ring diagrams  in the high-temperature regime. Demanding a second order phase transition as expected in the chiral limit at low baryon chemical potential, that the curvature of the critical line matches the one obtained in lattice simulations and that the value of the speed of sound  at high temperature and zero density has its measured value, we constrain the couplings and other free parameters of the model.  We set restrictions to locate the position of the Critical End Point in the phase diagram from a fast drop-off of the critical line reminiscent from the behavior of the speed of sound near criticality. Our results set tight constrains in parameter space for the model to exhibit realistic features.
\end{abstract}

\maketitle


As it is nowadays accepted, around one microsecond after the Big Bang, hadronic matter underwent a phase transition in which quarks and gluons in the primordial soup (the quark-gluon plasma or QGP for short) decoupled and hadronized~\cite{RAFELSKI2013155}. Traits of breaking and restoration of chiral symmetry are vital to understand the nature of strongly coupled matter. The conditions to reproduce such phase transition are met in relativistic heavy ion collision experiments in facilities like RHIC and LHC at low baryon densities and high temperatures, as recently reported in Ref.~\cite{arslandok2023hot}. In this case, the phase transition is a smooth and continuous cross-over, as suggested by lattice QCD (LQCD) simulations~\cite{Aoki2006,PhysRevD.74.054507,PhysRevLett.113.082001,PhysRevLett.113.152002,201915,PhysRevLett.125.052001,guenther2021overview} and confirmed in STAR~\cite{ADAMS2005102} and PHENIX~\cite{ADCOX2005184} experiments at RHIC for the first time and later refined at LHC. On the other hand, at extremely high densities and low temperatures, conditions met, for instance, in compact star environments, the transition is expected to be a discontinuous first order transition~\cite{ALFORD1998247,PhysRevLett.81.53,Krishna,Mark,schaefer2003quark,RISCHKE2004197}. For higher densities and lower temperatures as compared to those of existing experiments, other facilities like NICA~\cite{MPD:2022qhn,Kolesnikov_2020} and FAIR~\cite{Durante_2019} are under construction and expected to start running within a few years time. One ambitions scientific goal of these experiments is to locate the so-called Critical End Point (CEP), were the continuous and discontinuous phase transitions met as reported in the recent Ref.~\cite{vovchenko2023qcd}. LQCD simulations in these regimes are difficult to control due to the sign problem and other first-principle functional methods still demand modelling to achieve reliable predictions as shown in the state-of-the-art reports in Refs.~\cite{PhysRevD.101.054032,PhysRevD.104.054022,fu2023ripples}. Thus, a very good chance to address the behavior of hadronic matter in these circumstances can be obtained from effective model calculations.

One of such examples, the Linear Sigma Model with quarks (LSMq), has been successfully used to reproduce the chiral transition of strongly interacting matter under extreme conditions by appropriately taking into account the plasma screening effects in the dynamics near the transition~\cite{PhysRevC.79.015202,PhysRevD.93.114014,PhysRevD.68.016003,Ayala:2015hba,Ayala:2017dnn,Ayala:2017gek,Ayala:2017ucc,Ayala:2019skg,Ayala:2021tkm,Ayala:2023cnt,Mao_2010}. The behavior of the effective potential including ring diagrams~\cite{Dolan:1973qd} with respect to order parameter after spontaneous symmetry breaking is often employed to locate the CEP in thermodynamical parameter space. A favorite strategy to this end is to derive the statistical properties of the system from the fluctuations of the order parameter with respect to parameters such as energy of the collision, temperature and baryon chemical potential~\cite{Ayala:2021tkm}. Numerous studies have employed the linear sigma model or other effective models to analyze various thermal properties of the medium, such as the speed of sound, bulk viscosity, transport coefficients, among others~\cite{ayala2023describing,PhysRevD.109.076006,PhysRevD.86.074021,PhysRevC.102.034906,PhysRevD.107.096017,Bandyopadhyay2023}. In this letter, we focus on the behavior of transport quantities derived from the thermodynamics of the system near the transition. In particular, we consider the speed of sound for homentropic, $c_s$, isobaric $c_\rho$, and isentropico $c_{s/\rho}$ processes. It is expected  that exactly at the CEP, $c_{s/\rho}$ vanishes and that would be the smoking gun to locate this point in the phase diagram. 
To set constraints on the values of the couplings of the LSMq, we demand that $c_{s/\rho}\simeq c_\rho$ approaches the lattice extracted value at $\mu_B=0$~\cite{HotQCD:2014kol} and see how far it departures from zero as we permit the temperature and chemical potential to evolve in the phase diagram. 

We proceed from the Lagrangian of the LSMq, which reads 
\begin{eqnarray}
{\cal L}&=& \bar\psi \left(i {\not \! \partial} +g\sigma -i g \gamma^5 \vec\tau\cdot\vec\pi  \right)\psi +\frac12 \left( \partial_\mu \sigma\right)^2+\frac12 \left(\partial_\mu \vec \pi \right)^2\nonumber\\
&&+\frac{a^2}{2}\left(\sigma^{2}+\vec\pi^2 \right) -\frac{\lambda}{4}\left(\sigma^{2}+\vec\pi^2 \right)^2,\label{eq:lagsym}
\end{eqnarray}
where $\sigma$ represents the sigma boson, $\vec \pi$ is the pion triplet, with the neutral pion $\pi^0=\pi_3$ and the charged pions $\pi^\pm=(\pi_1 \mp \pi_2)/\sqrt{2}$, and $\psi$ is an isospin $SU(2)$ doublet representing the $u$ and $d$ quarks $\vec \tau$ are the Pauli matrices in isospin space. Mesons in this model are coupled with a four body contact interaction with strength $\lambda$ and with a three-point vertex with quarks with coupling strength $g$. Moreover, $a$ is a dimensionful parameter with mass units 1. Notice that when quarks remain massless, the Lagrangian~\eqref{eq:lagsym} is invariant under chiral transformations. This Lagrangian provides an effective description of the QCD phase diagram in terms of the chiral phase transition. Working in the strict chiral limit, to allow for a spontaneous symmetry breaking, we let the $\sigma$ field to develop a
vacuum expectation value $v$, namely $\sigma \rightarrow \sigma + v$, this vacuum expectation value is identified as the order parameter of the theory. After this shift, the Lagrangian can be rewritten as
\begin{eqnarray}
{\cal L}&=&\frac12(\partial_\mu \sigma)^2+\frac12(\partial_\mu \vec \pi)^2-\frac{1}{2}\left(3\lambda v^2 - a^2\right)\sigma^2\nonumber\\
&&-\frac{1}{2}\left(\lambda v^2 - a^2\right)\pi^2+\frac{a^2}{2}v^2-\frac{\lambda}{4}v^4\nonumber\\
&&+i \bar\psi \gamma^\mu\partial_\mu \psi +g v \bar\psi\psi +{\cal L}_I^b+{\cal L}_I^f ,
\label{eq:lagbrk}
\end{eqnarray}
in terms of ${\cal L}_I^b$ and ${\cal L}_I^f$, which are, respectively, given by
\begin{eqnarray}
    {\cal L}_I^b&=&-\frac{1}{4}\left[\left(\sigma^2+(\pi^0)^2\right)^2+4\pi^+\pi^-\left(\sigma^2+(\pi^0)^2+\pi^+\pi^-\right)\right] \nonumber\\
    {\cal L}_I^f&=& -g\bar{\psi}\left(\sigma+ i\gamma_5 \cdot\mathbf{\pi}\right)
\label{LfLb}
\end{eqnarray}
where the terms in Eq.~\eqref{LfLb} describe the interactions among the fields $\sigma$, $\mathbf{\pi}$ and $\psi$, after symmetry breaking. From Eq.~\eqref{eq:lagbrk} notice that the $\sigma$, the three pions and the quarks have masses given by
\begin{eqnarray}
    m_\sigma^2 &=& 3\lambda v^2 - a^2 \nonumber\\
    m_\pi^2 &=& \lambda v^2 - a^2 \nonumber\\
    m_f &=& gv 
\label{masses}
\end{eqnarray}

Note that in Eq.~\eqref{masses}, the squares of the tree-level bosons masses can either disappear or even turn negative as the continuous order parameter $v$ traverses its range. The physical masses, comprising their thermal components, are determined when $v$ adopts the value obtained from minimizing the effective potential. While our focus in this study has not been explicitly directed towards elucidating the behavior of these thermal masses, our findings align seamlessly with those of prior research, such as the study referenced in~\cite{PhysRevC.64.045202}, when accounting for the thermal contributions to these masses through their self-energies (refer to Eq.~\eqref{selfenergy} below), albeit in our case, under the high temperature regime.

The effective potential associated to \eqref{eq:lagbrk} at high temperature, up to order ring, reads
\begin{eqnarray}
	V_{\rm eff}(v)&=&\sum_{b=\pi^{\pm},\pi^0,\sigma}\Bigg\{-\frac{T^4\pi^2}{90}+\frac{T^2m_b^2}{24}-T\frac{\left(m_b^2+\Pi_b\right)^{3/2}}{12\pi}\nonumber\\
 &&-\frac{m_b^4}{64\pi^2} \left[\mathrm{ln}\left(\frac{\mu^2}{\left(4\pi T\right)^2}\right)+2\gamma_E\right] \Bigg\} \nonumber \\ 
 &&   + N_c N_f \Bigg\{ \frac{m^4_f}{16\pi^2} \Bigg[ \mathrm{ln}\left(\frac{\mu^2}{T^2}\right)  -\psi^0\left(\frac{1}{2}+\frac{i\mu_q}{2\pi T}\right)\nonumber\\
 &&-\psi^0\left(\frac{1}{2}-\frac{i\mu_q}{2\pi T}\right) +\psi^0\left(\frac{3}{2}\right)  \nonumber\\
 &&-2(1+\mathrm{ln}(2\pi)) +\gamma_E \Bigg]\nonumber \\
 &&  -\frac{m^2_fT^2}{2\pi^2}\left[\mathrm{Li}_2\left(-e^{-\frac{\mu_q}{T}}\right)+\mathrm{Li}_2\left(e^{-\frac{\mu_q}{T}}\right)\right] \nonumber  \\ 
 &&  +\frac{T^4}{\pi^2}\Bigg[\mathrm{Li}_4\left(-e^{-\frac{\mu_q}{T}}\right)+\mathrm{Li}_4\left(e^{-\frac{\mu_q}{T}}\right)\Bigg] \Bigg\}\nonumber\\
 &&-\frac{a^2}{2}v^2+\frac{\lambda}{4}v^4,
\end{eqnarray}
where $\mathrm{Li}_n(x)$ is the polylogarithm function of order  $n$ and $\psi^0(x)$ is the polygamma function,  $\gamma_E \approx 0.57721$ is the Euler-Mascheroni constant, $\mu_q$ denotes the quark chemical potential, related to the baryon chemical potential simply as  $\mu_q=\mu_B/3$, $\mu$ is a scale parameter arising in the regularization scheme and that we fix to $\mu=500 \mathrm{MeV}$, $N_f$ and $N_c$ are te number of flavors and colors, respectively, which we fix as $N_f=2$ and $N_c=3$.  $\Pi_b$ is the boson self-energy including screening effects~\cite{Dolan:1973qd} that in the high temperature regime is given by
\begin{eqnarray}
	\Pi_b\equiv \Pi_\sigma&=&\Pi_{\pi^\pm}=\Pi_{\pi^0}=\nonumber\\
 &&\hspace{-2cm}\lambda\frac{T^2}{2}-N_fN_cg^2\frac{T^2}{\pi^2}\left[\mathrm{Li}_2\left(-e^{-\frac{\mu_q}{T}}\right)+\mathrm{Li}_2\left(e^{-\frac{\mu_q}{T}}\right)\right].
\label{selfenergy}
\end{eqnarray}
The order ring terms $\propto(m_b^2+\Pi_b)^{3/2}$ allow to incorporate the plasma screening effects necessary to get rid of the potentially dangerous pieces coming from linear or cubic powers of the boson mass, that could become
imaginary for certain values of $v$. 

We fix the couplings $g$ and $\lambda$ along with the scale $a$ as follows. Knowing that 
at high temperature, in the chiral limit the phase transition is of second order, 
\begin{equation}
	\left.\frac{\partial^2 V_{\rm eff}}{\partial v^2} \right| _{v=0}=0,
\end{equation}
we obtain the following condition
\begin{eqnarray}
a^2	&=& \lambda \Bigg(\frac{T_c^2}{12}-\frac{T_c}{4 \pi }\sqrt{\Pi_b-a^2}\nonumber\\
 &&\hspace{-3mm}+\frac{a^2}{16\pi^2} \left[\mathrm{ln} \left(\frac{\mu^2}{(4 \pi T_c)^2}\right)+2 \gamma_E \right]\Bigg) \nonumber \\ 
 &&\hspace{-3mm}-N_fN_cg^2\frac{T_c^2}{\pi^2}\left[\mathrm{Li}_2\left(-e^{-\frac{\mu_{qc}}{T_c}}\right)+\mathrm{Li}_2\left(e^{-\frac{\mu_{qc}}{T}}\right)\right], \label{Veff2}
\end{eqnarray}
that allows us to fix the scale parameter $a$ in terms of $g$ and $\lambda$ by taking into account the LQCD result for  $N_f=2$ and $\mu_{q}^c=0$, which states that the critical temperature for the chiral phase transition is $T_c=166 \ \mathrm{MeV}$ \cite{Gao:2020qsj}. 
Moreover, also from LQCD~\cite{Guenther:2020jwe}, we have that the transition curve is parameterized as
\begin{equation}\label{TF}
	\frac{T_c(\mu_q)}{T^0_c}=1-9 \kappa_2  \left(\frac{\mu_q}{T^0_c}\right)^2+81 \kappa_4 \left(\frac{\mu_q}{T^0_c}\right)^4,
\end{equation}
with $\kappa_2=0.0139(18)$ and $\kappa_4=0.0003951$~\cite{Guenther:2020jwe}. Varying $\lambda$ in a window of values permits to pick the best choice of $g$ to ensure the correct curvature in our sketch of the transition line. 
A third condition to fix the parameters of the model emerges by recalling that the pressure is given by $p=-V_{\rm eff}$ and the energy density is $\epsilon=-p+Ts+\mu_q\rho$, where quantities carry their usual meaning, such that the entropy density is
\begin{equation}
	s=\frac{\partial p}{\partial T}\Bigg|_{\mu_q} \hspace{10mm}\mathrm{and}\hspace{10mm} \rho=\frac{\partial p}{\partial \mu_q}\Bigg|_{T}.
\end{equation}
Thus, the speed of sound $c^2_x$ for different parameters $x$ can be found using the Jacobian method and suitable thermodynamical relations. In particular,
\begin{eqnarray}
	c^2_{\rho}&=&\frac{\partial p}{\partial\epsilon}\Bigg|_\rho=\frac{s\chi_{\mu_q\mu_q}-\rho\chi_{\mu_q T}}{T\left(\chi_{TT}\chi_{\mu_q\mu_q}-\chi^2_{\mu_q T}\right)},\label{cp}\\
	c^2_{s}&=&\frac{\partial p}{\partial\epsilon}\Bigg|_s=\frac{\rho_B\chi_{TT}-s\chi_{\mu_q T}}{\mu_q\left(\chi_{TT}\chi_{\mu_q\mu_q}-\chi^2_{\mu_q T}\right)},\label{cs} \\
	c^2_{s/\rho}&=&\frac{\partial p}{\partial\epsilon}\Bigg|_{\rho,s}=\frac{c^2_{\rho}Ts+c^2_s\mu_q\rho}{Ts+\mu_q\rho}, \label{csp}
\end{eqnarray} 
are, respectively, the isobaric, homentropic and isentropic speeds of sound. In the above expressions,
the second order susceptibility $\chi_{xy}$ is defined as
\begin{equation}
	\chi_{xy}=\frac{\partial^2 p}{\partial x \partial y},
\end{equation} 
knowing that $c_{s/\rho}^2\simeq0.16$~\cite{HotQCD:2014kol} for $\mu_{q}^c=0$ and $T_c=166 \ \mathrm{MeV}$, we reach to a third condition that must be satisfied by the tree parameters $a$, $g$ and $\lambda$. 

\begin{figure}[]
    \centering
    \includegraphics[width=0.9\columnwidth]{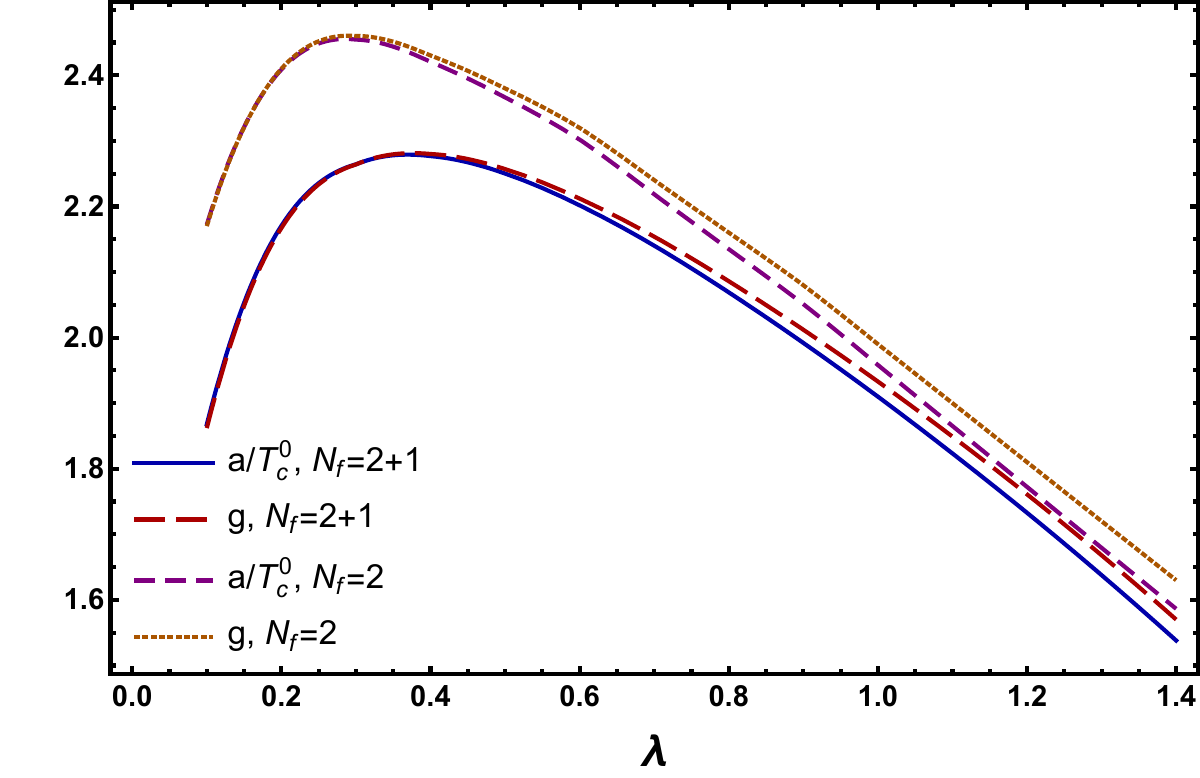}
    \caption{Parameters $g$ and $a/T_c^0$ as functions of $\lambda$. These are obtained by assuring that the transition for $\mu_q = 0$ occurs exactly at $T_c = 158 {\rm MeV}$, which corresponds in lattice simulations to $N_f=2+1$ and  at $T_c = 166 {\rm MeV}$ for $N_f=2$. The parameters $\kappa_2$ and $\kappa_4$ of the curvature for small values of $\mu_q$ coincide with the values of the Refs.~\cite{PhysRevLett.125.052001,Guenther:2020jwe}.}
    \label{ganda}
\end{figure}

\begin{figure}[]
    \centering
    \includegraphics[width=0.9\columnwidth]{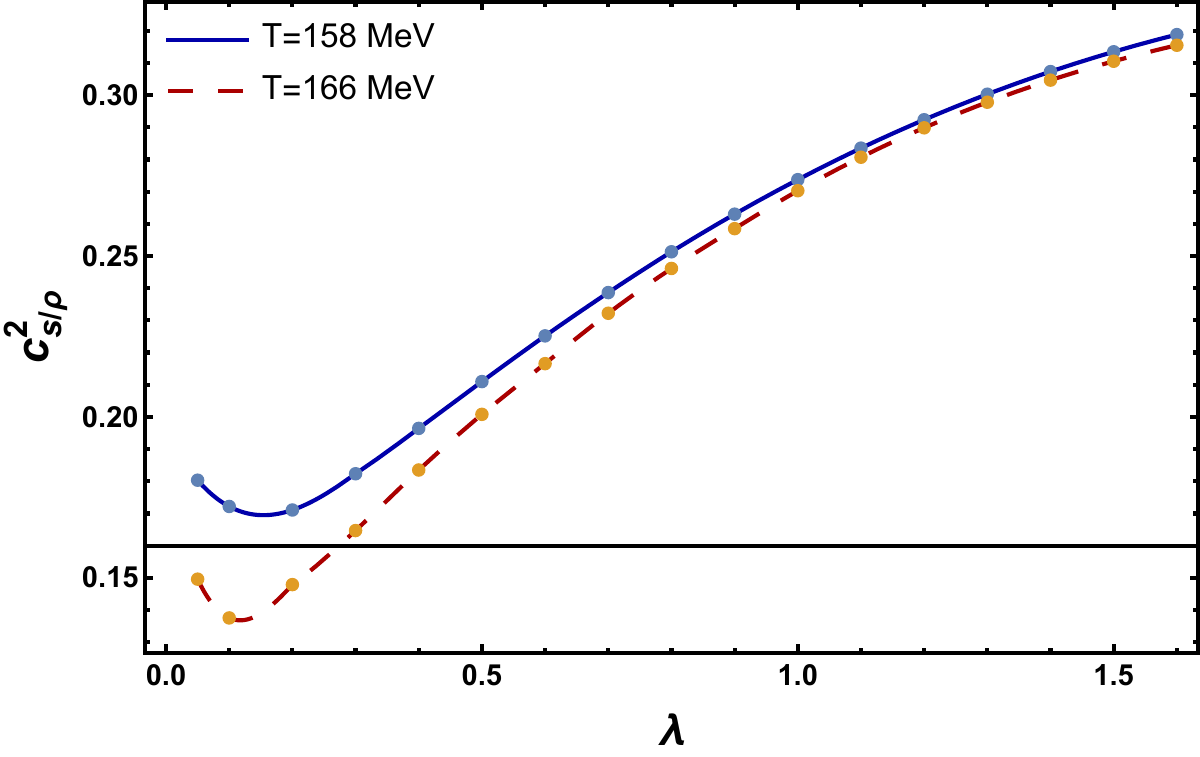}
    \caption{Speed of sound dependence on the quartic coupling $\lambda$ for $T_c = 158{\rm MeV}$ and $T_c = 166{\rm MeV}$, corresponding to $N_f=2+1$ and $N_f=2$ flavors, respectively. Straight line shows the value of $c_{s/\rho}^2\simeq0.16$ obtained from~\cite{HotQCD:2014kol}.}
    \label{ganda2}
\end{figure}

In Fig.~\ref{ganda} we display the sensitivity of $g$ and $a/T_c^2$ to variations of $\lambda$. We depict the dependence of these quantities assuming the critical temperature of $N_f=2$ and $N_f=2+1$ flavors in LQCD simulations. The traits of the dependence on the quartic coupling and the parameter $a/T_c^0$ are similar for both these simulations. Figure~\ref{ganda2} sketches the behavior of the speed of sound $c_{s/\rho}^2$ as a function of the same coupling. We observe that only for small values of $\lambda$ we can find values of the speed of sound compatible with measurements, signal that there should be a rapid variation in these parameters compared to the values in vacuum. Here we automatically observe that the constraint that the values of the parameters $g$, $\lambda$ and $a$ should be compatible with $c_{s/\rho}$ decrease drastically the space parameter region. 

\begin{figure}[t!]
    \centering
    \includegraphics[width=0.9\columnwidth]{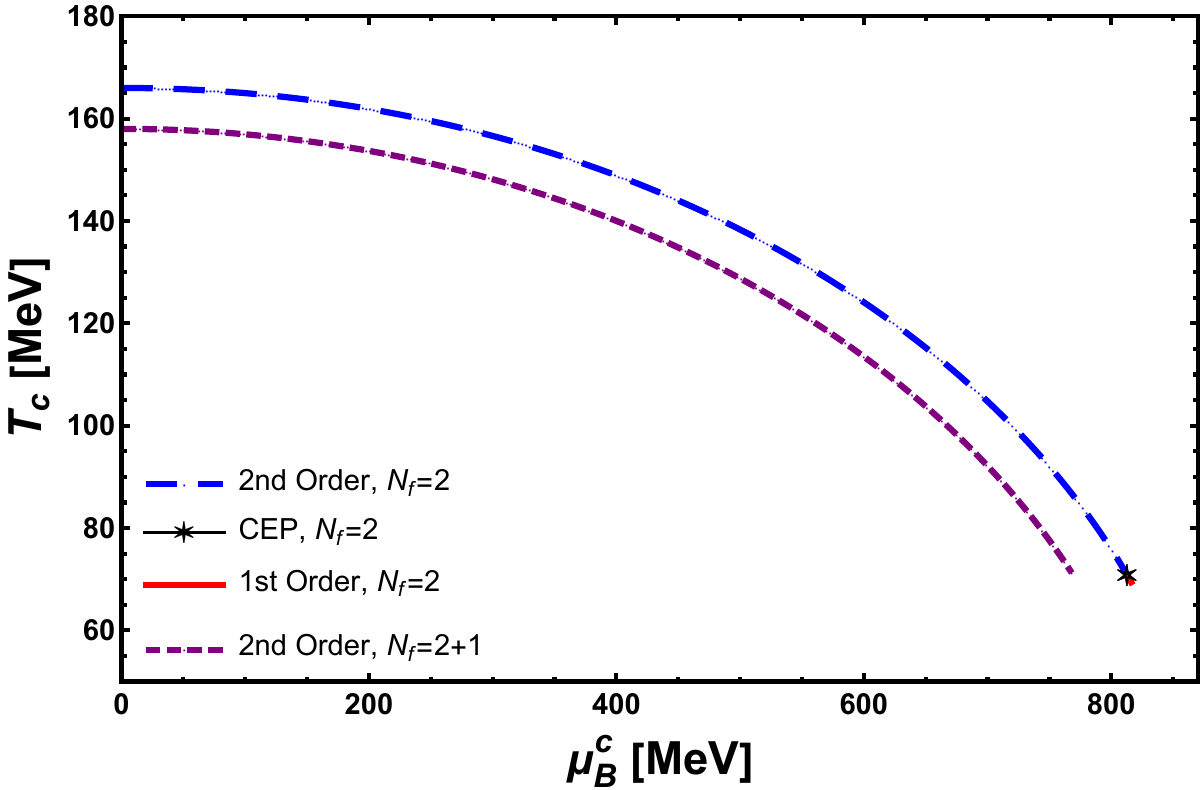}
    \caption{Examples of the effective phase diagram obtained with the parameters $a$, $\lambda$ and $g$ that ensure the value of $c_{s/\rho}^2 = 0.16$ at $T_c^0 = 166 {\rm MeV} $ and $T_c^0 = 158 {\rm MeV} $.}
    \label{pdiag}
\end{figure}

Examples of the shape of the phase diagram that we can draw from our perspective are depicted in Fig.~\ref{pdiag} for values corresponding to $T_c^0$ for $N_f=2$ and $N_f=2+1$ flavors in LQCD simulations. We select the speed of sound $c_{s/\rho}=0.16$ for this exercise. We notice that the first order transition is achieved for small values of temperature, $T_E\simeq 80{\rm MeV}$, and values of the baryon chemical potential $\mu_{B}^{E}\simeq 800{\rm MeV}$ for $N_f=2$, but for $N_f=2+1$ the CEP cannot be located, as the high-temperature assumption is no longer reliable.

\begin{figure}[h!]
    \centering
    \includegraphics[width=0.9\columnwidth]{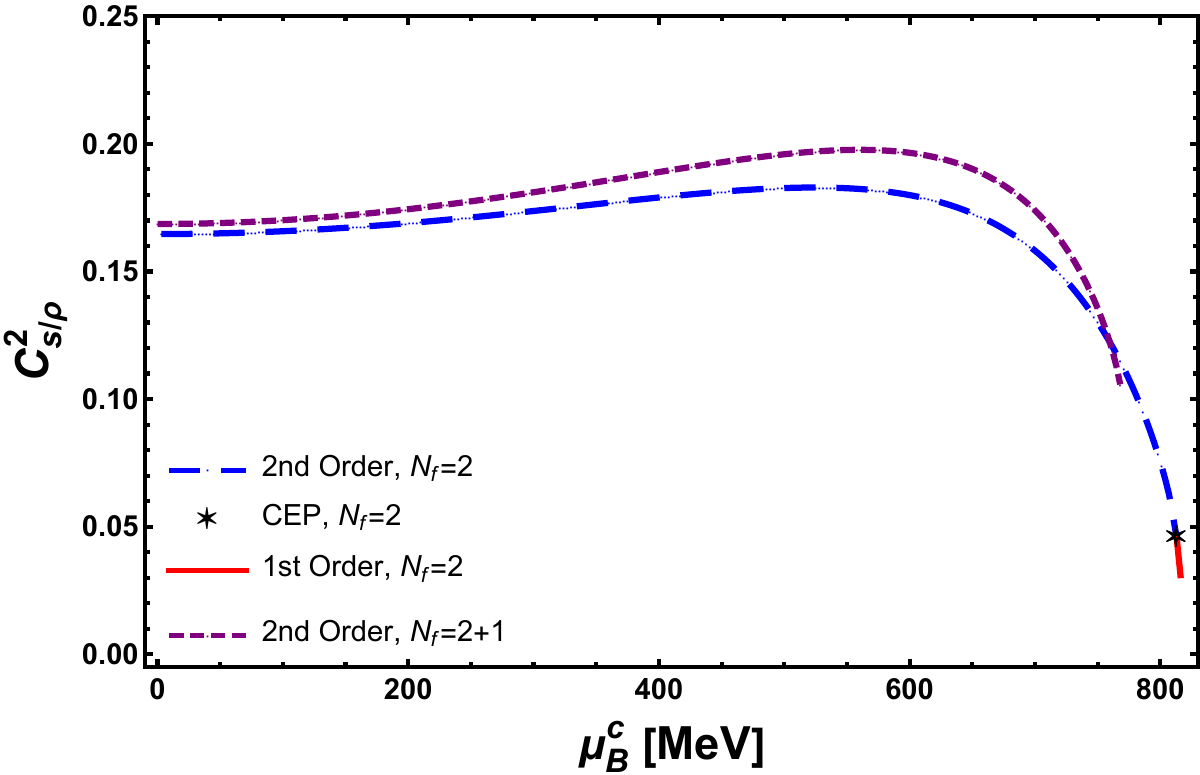}
    \caption{Behavior of the speed of sound  for different critical $\mu_B^c$ and its corresponding $T_c$. For $N_f=2$, we distinguish the order of the transition and show the CEP. We can notice how there is a significant decrease in the value of $c_{s/\rho}^2$ as we approach to CEP. No evidence of the existence of a CEP is observed for $N_f=2+1$ flavors, for which the high-$T$ approximation ceases to be valid.}
    \label{sos}
\end{figure}

In Fig.~\ref{sos} we depict the behavior of  $c_{s/\rho}^2$ for different critical values $\mu_B^c$ and the corresponding critical temperatures $T_c$. Again, different curves are for the parameters of $N_f=2$ and $N_f=2+1$ in LQCD.  For $N_f=2$, the CEP can be located and a sharp fall-off of $c_{s/\rho}^2$ is noticeable which can be used as a signal to detect its position. This is not the case of $N_f=2+1$, where no evidence of first order transition can be found because the high-$T$ approximations breaks before reaching the CEP.


In this letter we re-examined the LSMq in connection with the physics of the QCD phase diagram and the existence of the CEP. Although it is often desirable to include the effects of strangeness and nuclear matter and their role in the location of the CEP, it is well established that neither one of these  have a dominant thermodynamic contribution near criticality. Thus, we decided simply omit them from our considerations. Rather, we have put forward a strategy to fix the couplings $g$ and $\lambda$ as well as the parameter $a$ in the Lagrangian~\eqref{eq:lagsym} from three requirements: (i) the condition of continuous phase transitions at high temperature, \eqref{Veff2}, (ii) the curvature of the critical line in the phase diagram according to LQCD, \eqref{TF}, and (iii) demanding that the speed of sound $c_{s/\rho}^2$ corresponds to the measured value. We consider that such a strategy allows to further constrain the free parameter space and discard some approximations in the need for more educated analysis of the phase diagram through effective models of this kind. As compared with a previous analysis by some of us regarding the collision energy dependence in the position of the CEP~\cite{Ayala:2021tkm}, we have narrowed down the range of values of $\lambda$ to a single point. We stress that the attainable values of $c_{s/\rho}$ in that work are significantly separated from the measured value. In this regard, our demand that the parameter choice renders the measured value of the speed of sound is in fact a tight constraint that any treatment of the effective potential of the LSMq should fulfill.

It is interesting that our fixing parameter scheme also calls for the need to abandon the high-temperature approximation for the effective potential and perhaps next-to-leading terms or low temperature approximations are better suited for sketching the phase diagram if the CEP should exist at all~\cite{Ayala:2017gek,Ayala:2017ucc}. Otherwise, the LSMq is uncapable of finding the CEP in the phase diagram.

SHO and AR acknowledge support from  Consejo Nacional de Humanidades, Ciencia y Tecnología (México) under grant CF-2023-G-433 as well as Consejo de la Investigación Científica (UMSNH, México) under project 18371. RMvD acknowledge support from the project CONAHCyT (México) CF-428214.

\bibliography{Referencias}

\end{document}